\documentclass[prb,aps,twocolumn,superscriptaddress,showpacs]{revtex4}

\usepackage{epsfig}

\newcommand{\be}{\begin{equation}}
\newcommand{\ee}{\end{equation}}
\newcommand{\bea}{\begin{eqnarray}}
\newcommand{\eea}{\end{eqnarray}}
\newcommand{\ham}{{\cal H}}

\newcommand{\reff}{{\mbox{eff}}}

\newcommand{\p}{\partial}
\newcommand{\s}{\sigma}

\newcommand{\la}{\langle}
\newcommand{\ra}{\rangle}
\newcommand{\rd}{\mbox{d}}
\newcommand{\ri}{\text{i}}

\newcommand{\up}{\uparrow}
\newcommand{\down}{\downarrow}
\def\nn{\nonumber\\}

\begin{document}

\title{Interaction induced dimerization in zigzag single wall carbon nanotubes}

\author{Sam T. Carr}
\affiliation{School of Physics and Astronomy, University of Birmingham, Birmingham B15 2TT, United Kingdom}
\author{Alexander O. Gogolin}
\affiliation{Department of Mathematics, Imperial College, 180 Queen's Gate, London SW7 2BZ, United Kingdom}
\author{Alexander A. Nersesyan}
\affiliation{The Abdus Salam International Centre for Theoretical Physics, 34100, Trieste, Italy}
\affiliation{The Andronikashvili Institute of Physics, Tamarashvili 6, 0177, Tbilisi, Georgia}

\date{\today}


\begin{abstract}

We derive a low-energy effective model of metallic zigzag carbon nanotubes at half filling.  
We show that there are three important features characterizing the low-energy properties of these systems:
the long-range Coulomb interaction, 
umklapp scattering and an explicit dimerization generated by interactions.
The ratio of the dimerization induced gap and the Mott gap induced by the umklapp 
interactions is dependent on the radius of the nanotube and can drive the system through a quantum 
phase transition with $SU(2)_1$ quantum symmetry.  We consider the physical properties of the phases 
on either side of this transition
 which should be relevant for realistic nanotubes.
\end{abstract}

\maketitle

\section{Introduction}

Since their discovery,\cite{i91} carbon nanotubes have 
attracted a great amount of attention, both theoretically and experimentally.\cite{d99} 
It is widely appreciated that
single-wall  nanotubes (SWNT)
constitute almost ideal systems where peculiar effects specific to strong
correlations in one dimension (1D) can be observed.
They are structurally
1D objects built by wrapping a sheet of graphene into a cylinder.
The type of wrapping is 
characterized by the superlattice vector $(n,m)$. 
From the point of view of strong correlations metallic
nanotubes are of particular interest. 
In the absence of
interactions between the electrons,\cite{saito} the band structure is indeed metallic for 
all $(n,n)$ (or {\it armchair}) nanotubes,
as well as for the $(n,-n)$ (or {\it zigzag}) nanotubes with $n$ being multiple of 3.
In all other cases there is band gap which at large $n$ is of the order of $(10/n) eV$.

The low-energy effective theory for correlated metallic SWNTs away from half-filling
\cite{eg98}
shows that the
long-range Coulomb interaction converts the nanotubes into Tomonaga-Luttinger liquids
implying various scaling laws for conductance, which have since been verified
experimentally.\cite{bc+99} At half-filling the effects of strong
correlations become even more pronounced 
in the presence of the poorly screened Coulomb interaction. The latter makes the umklapp 
scattering processes strongly relevant giving rise to sizable Mott gaps.\cite{yo99}
The underlying strong-coupling phases for armchair nanotubes were recently classified in 
Ref. \onlinecite{nt03}.

In this paper, we study the effects of the long-range Coulomb interaction in half-filled
zigzag SWNTs ($n$ divisible by $3$).  The new feature that makes this case 
qualitatively
different from the armchair
nanotubes is an explicit dimerization that originates from the nearest-neighbor interaction 
and gives rise to a single-particle gap of the order of $1/n$.
Notice that curvature effects,\cite{km97} 
which are known to lead to much smaller band gaps, $\sim 1/n^2$, 
can 
therefore be safely neglected. 
We will explain how the dimerization comes about, discuss its interplay with the Mott gaps
and show that this competition can result in a quantum phase transition 
similar to the one studied previously in the context of dimerized spin ladders.\cite{mss96,wn00}
Making reasonable approximations for the relative strengths of different interaction terms for a zigzag SWNT shows that the system can occur very close to
the quantum criticality, and may even be tuned to reach it by changing the radius of the nanotube; however a realistic estimation of the parameters by either experiment or numerical techniques is beyond the scope of this paper.

Below we will follow the same strategy as in the armchair case \cite{nt03}
paying particular attention to the differences.  
The structure of the paper is as follows: First we consider the band-structure of the zigzag 
SWNT within a tight-binding model and show that,
in the low-energy limit, the latter is equivalent to an effective two-chain model.
We then very carefully consider the structure of the interaction, 
taking into account both the long-range tail of the unscreened Coulomb interaction and the details of the 
short-range component of the interaction which depend in an important way on the lattice.  We then solve 
this model by using bosonization, employing the adiabatic approximation
to treat neutral collective excitations, and then concentrating on energies well below
the charge gap,  by refermionizing the 
remaining theory.  This allows us to identify the nature of the phase transition generated by the 
dimerization term, and extract
some physical properties of such nanotubes.


\section{Mapping onto low energy effective theory}


In this section, we single out those two bands that cross the Fermi level in  an undoped 
isolated zigzag SWNT, which allows us to describe the relevant part of the spectrum
in terms of an equivalent two-chain fermionic model.
We carefully analyze the details of the lattice structure that determine
the peculiarities of the interaction terms.  We then pass to the continuum limit 
and employ the bosonization 
technique to derive the low-energy effective field theory for a metallic zigzag SWNT.

\subsection{Kinetic term and two chain model in zigzag nanotubes}

\begin{figure}
\begin{center}
\epsfig{file=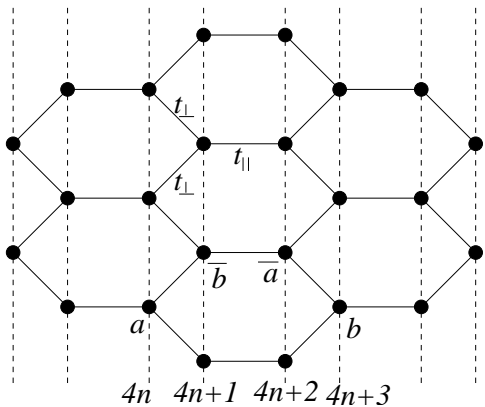,width=2.5in}
\caption{Real space structure of a zigzag nanotube.  
In this paper
the chain direction is chosen to be
$x$, 
meaning that the axis of the graphite sheet have been 
rotated by $90^o$ 
as compared to the references that deal with the armchair case. }\label{zz1}
\end{center}
\end{figure}

We begin with a tight-binding model on the honeycomb graphite   
lattice shown in Fig.1.  The structure consists of an underlying triangular lattice with two inequivalent
 Carbon positions, labeled by $a$ and $b$.
It proves convenient to double the unit cell such that the new cell
contains four atoms, $a,\bar{b},\bar{a},b$.  In this picture the unit cell is rectangular, so the two momenta, $k_x$ and $k_y$, can be considered as independent.
Wrapping a $(N \times n)$ graphite sheet into a $(n, -n)$ zigzag nanotube quantizes the
electron momentum in the 
direction of the superlattice vector ${\bf C} = n ({\bf a}_1 - {\bf a}_2)$, i.e. the $y$ direction in our Cartesian axis.  Carrying out
a partial Fourier transform in this direction gives a Hamiltonian describing
$n$ 1D bands (see e.g. Ref. \onlinecite{lin98}),
\be
H_0 = -\sum_q \sum_{l=1}^{4N} t_q(l) \left[ c_q^\dagger (l) c_q (l+1) + h.c. \right].
\label{1d-ham}
\ee
Here $l$ labels the sites of an effective 1D lattice along the tube's axis, the lattice
spacing being $x_{l+1} - x_l \equiv b = \sqrt{3}/4$,
$q$ is the band index taking $n$ integer values ($|q| < n/2$), and 
$t_q(l) = t_0 (q) + (-1)^l \Delta (q)$ is an alternating hopping amplitude whose uniform and staggered
parts are given by
\bea
t_0(q) &=& \frac{1}{2}\left[t_{\parallel} + 2t_{\perp} \cos(\pi q/n)\right],\label{uniform}\\
\Delta(q) &=&  \frac{1}{2}\left[t_{\parallel} - 2t_{\perp} \cos(\pi q/n)\right].\label{dimer}
\eea

Thus the spectrum is generically gapped, the gap of the $q$th band being determined by the dimerization
amplitude (\ref{dimer}). 
However, for two bands labeled by $q = \pm Q = \pm n/3$ ($n=3m$) 
the gap is minimal, proportional to the difference
$t_{\parallel} - t_{\perp}$, and is solely due to curvature effects.
This difference was shown\cite{km97} to be of the order $1/n^2$ and is much smaller
than the gaps of all the other bands. For this reason, and also in view of the fact that
the interactions are of the order $1/n$ or larger, one can set
$t_{\parallel} = t_{\perp}$ and thus
make the lowest-energy part of the spectrum described in terms of two 
decoupled, translationally invariant chains with a simple half-filled cosine band, via the mapping
\bea
b(m_x,m_y) &\rightarrow& \frac{1}{\sqrt{n}}\left[
e^{\ri Qy} c_+ (4m_x) + e^{-\ri Qy} c_- (4m_x)\right] \nn
a(m_x,m_y) &\rightarrow& \frac{1}{\sqrt{n}}\left[
e^{\ri Qy} c_+ (4m_x + 1) + e^{-\ri Qy} c_- (4m_x +1)\right] \nn
\overline{b}(m_x,m_y) &\rightarrow& \frac{1}{\sqrt{n}}\left[
e^{\ri Q(y+1/2)} c_+ (4m_x + 2) \right. \nn
 && \;\;\;\;\;\; \left. + e^{-\ri Q(y+1/2)} c_- (4m_x +2)\right] \nn
\overline{a}(m_x,m_y) &\rightarrow& \frac{1}{\sqrt{n}}\left[
e^{\ri Q(y+1/2)} c_+ (4m_x + 3) \right. \nn
&& \;\;\;\;\;\; \left. + e^{-\ri Q(y+1/2)} c_- (4m_x +3)\right]. \label{twochainmapping}
\eea
This picture is in contrast with the case of armchair nanotubes where the kinetic part of the low-energy 
theory is that of two strongly coupled chains (ladder).
Even though curvature effects can be neglected, 
the dimerization $\Delta$ should be retained in 
the Hamiltonian.  This follows from the fact that 
interactions generate an identical term in 
the low-energy effective theory, but with  a much larger amplitude, $\sim 1/n$.  The next 
two sections will demonstrate this phenomena, which is one of the central results of this paper.

\subsection{Interactions}

There are two important contributions to the interactions.
Firstly, 
in an isolated nanotube, the
Coulomb interaction is unscreened, so one must take into account the effects of its long-range tail.  Secondly,
at half-filling,
Umklapp processes originating from the short-range part
of the Coulomb interaction play a crucial role.
Below we consider these two contributions separately.

The long-range part of the interaction is insensitive to the details of the lattice and is simply 
given by 
\be
H_{\rm Coul} = \sum_{ll'} n(l) U(l-l') n(l'),
\ee
where $n(l) = \sum_{\mu\sigma} c^\dagger_{\mu\sigma} c_{\mu\sigma}$ is the total density on lattice site $l$ in the
two chain mapping, and $U(x) \sim e^2/|x|$.

\begin{figure}
\begin{center}
\epsfig{file=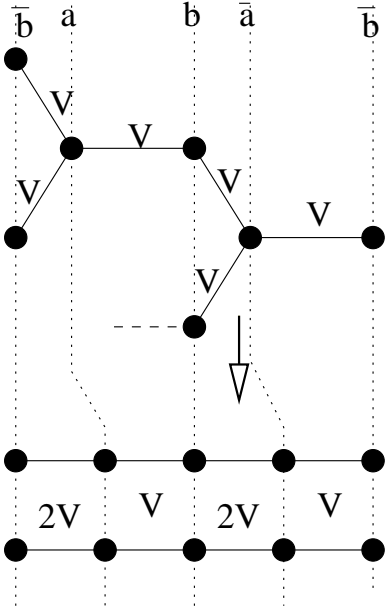,width=2.0in}
\caption{A cartoon representation of the origin of the staggered interaction}\label{stag}
\end{center}
\end{figure}

The effects of the short-range part of the interaction are best elucidated 
by considering the minimal model that
captures all essential physics of the problem.
It turns out that a two-parameter model with on-site ($U$) and nearest-neighbour ($V$) 
couplings 
fully represents the most general structure of local interaction in the low-energy limit,
\bea
H_1 &=& U \sum_{m_x,m_y} n_\uparrow(m_x,m_y)n_\downarrow (m_x,m_y) \nn
&+& V \sum_{<n.n.>} n(m_x,m_y)n(m'_x,m'_y).
\eea
Here $m_x,m_y$ label the carbon atoms on the original 2D lattice, 
the summation in the second term goes over nearest neighbours,
and $n(m_x,m_y)=n_\uparrow(m_x,m_y)+n_\downarrow(n_x,n_y)$ is the electron density on 
site $(m_x,m_y)$.
When the $V$-term is projected onto the two low-energy chains, we see an interesting 
effect as illustrated in 
Fig. \ref{stag}: each atom $a$ ($\overline{a}$) interacts with one atom $b$ ($\overline{b}$) and with
two atoms $\overline{b}$ ($b$).
In other words, when mapping onto the two-chain low-energy sector,
the interaction $V_{ab}=V_{\overline{a}\overline{b}} \ne V_{a\overline{b}} = V_{\overline{a}b}$.  
This means that in the low-energy theory the $V$ interaction 
will contain not only a uniform part but also a staggered part.
The latter is entirely due to the way we break 
the $C_3$ symmetry of the 2D lattice by wrapping it to form a zigzag nanotube; the effect 
is not present in the armchair case.  Carrying out the mapping (\ref{twochainmapping}) to the effective 
two-chain model, with index $\mu = \pm$ labeling the two chains, yields three terms, 
\bea
H_U &=& \frac{U}{n}\sum_{l\mu} n_{\mu\uparrow}(l) n_{\mu\downarrow}(l)
+ \frac{U}{n} \sum_{l\mu} c_{\mu\uparrow}^\dagger c_{-\mu\uparrow} c_{-\mu\downarrow}^\dagger c_{\mu\downarrow} \nn
H_V &=& \frac{3V}{2n} \sum_{l} n(l) n(l+1) \nn
H_s &=& -\frac{V}{2n} \sum_l (-1)^l n(l) n(l+1) \nn
&& \,\, + \frac{V}{n} \sum_{l\mu\sigma\sigma'}(-1)^l c_{\mu\sigma}^\dagger c_{-\mu\sigma} c_{-\mu\sigma'}^\dagger c_{\mu\sigma'},\label{interaction}
\eea
where $H_U$ is the contribution from the on-site interaction, $H_V$ is the smooth contribution from the nearest neighbor interaction, and $H_s$ is the staggered part of the nearest neighbor interaction.
We now proceed to study the ground state and elementary excitations of such an interacting 
two chain model by employing the bosonization technique'.


\subsection{Chiral Decomposition and Bosonization}

Passing to the continuum limit, we adopt the standard description in terms of chiral 
(right/left) fermions.  
At half-filling  $k_F=\pm \pi/2b$, so that 
\be
c_{\mu\sigma} (l) \rightarrow \sqrt{b} \left[ \ri^l
R_{\mu\sigma} (x) + (-\ri)^l
L_{\mu\sigma} (x)\right],
\ee
where 
$\s = \up,\down$ is the spin index.
This yields $H_0 = \int \rd x~ {\cal H}_0 (x)$, where
\bea
&&{\cal H}_0 =  -\ri v_F \left( R^\dagger_{\mu\sigma}\partial_x R_{\mu\sigma}
- L^\dagger_{\mu\sigma}\partial_x L_{\mu\sigma} \right)
+ 2 g_\Delta {\cal O}_{\dim}, \nonumber\\
&&{\cal O}_{\dim}
= \ri \left(  R^\dagger_{\mu\sigma} L_{\mu\sigma} - L^\dagger_{\mu\sigma} R_{\mu\sigma} \right).
\label{dim-op}
\eea
To treat interactions nonperturbatively, 
we employ Abelian bosonization based on the correspondence
$
R(L)_{\mu\s} \to (2\pi\alpha)^{-1/2} 
\exp \left[-i \sqrt{\pi}\left( \Phi_{\mu\s} \mp \Theta_{\mu\s} \right)  \right],
$
where $\Phi_{\mu\s}$ and $\Theta_{\mu\s}$ are a pair of mutually dual scalar fields,
and $\alpha$ is a short-distance
cutoff of the bosonic theory.\cite{footnote}
Following Ref. \onlinecite{ek92}, we pass to linear combinations of the bosonic fields describing
the total and relative charge and spin excitations,
\bea
\Phi^{\pm}_c &=& \frac{1}{2}\left( \Phi_{+\up} + \Phi_{+\down} 
\pm \Phi_{-\up} \pm \Phi_{-\down} \right),
\nonumber\\
\Phi^{\pm}_s &=& \frac{1}{2} \left( \Phi_{+\up} - \Phi_{+\down} 
\pm \Phi_{-\up} \mp \Phi_{-\down} \right).
\nonumber
\eea
The kinetic energy of chiral Fermions 
then becomes the sum of four Gaussian models ($a = c^\pm,s^\pm$),
\be
\ham_{\rm kin} = (v_F / 2)\sum_a \left[ \left( \partial_x \Theta_a \right)^2 + 
\left( \partial_x \Phi_a \right)^2 \right],
\ee
while the dimerization operator takes the form:
\bea
&& {\cal O}_{\rm dim} = (4/\pi\alpha) \nonumber\\
&& \times [ \cos(\sqrt{\pi}\Phi_c^+) \cos(\sqrt{\pi}\Phi_c^-) \cos(\sqrt{\pi}\Phi_s^-) 
\cos(\sqrt{\pi}\Phi_s^+)  \nonumber\\
&& +  \sin(\sqrt{\pi}\Phi_c^+) \sin(\sqrt{\pi}\Phi_c^-) 
\sin(\sqrt{\pi}\Phi_s^-) \sin(\sqrt{\pi}\Phi_s^+)]. \nn
&&
\label{dimop2}
\eea

As in the armchair case, 
the long-range part of the unscreened Coulomb 
interaction only involves 
the total charge field
\be
H_{\rm Coul} = \frac{2e^2}{\pi} \int \rd x \int \rd y 
\frac{ \p_x \Phi^+ _c (x)~ \p_y \Phi^+ _c (y)}{|x-y|},
\label{lr-coul}
\ee
and greatly reduces the scaling dimension of the operator $\cos(\sqrt{4\pi}\Phi_c^+)$.
At half filling, Umklapp processes contain this operator, and therefore become strongly relevant giving rise to 
a significant increase of all the gaps induced by interaction. \cite{yo99,nt03} 
\begin{figure}
\begin{center}
\epsfig{file=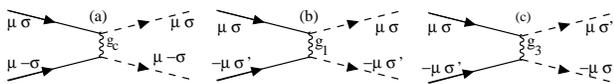,width=3.2in}
\caption{The umklapp interaction g-ology: (a) in-chain, (b) inter-chain, no spin exchange, (c) inter-chain, 
spin exchange. 
The solid and dashed lines describe particles with opposite chiralities.}\label{intg}
\end{center}
\end{figure}

Turning to the uniform part of the short-range interaction (\ref{interaction}), we single out
contributions of umklapp processes, all of them proportional to the operator $\cos(\sqrt{4\pi}\Phi_c^+)$. 
This selection can be done 
by directly bosonizing (\ref{interaction}), and can also be understood 
physically by using the g-ology approach (see Fig. \ref{intg}).  
The result is:
\bea
\ham_{\rm umkl} &=& - \frac{b}{(\pi\alpha)^2}
\cos(\sqrt{4\pi}\Phi_c^+) \left\{
g_c \cos(\sqrt{4\pi}\Phi_c^-) \right. \nn
&+&  (g_3-g_1)  \cos(\sqrt{4\pi}\Phi_s^+)
+ g_1  \cos(\sqrt{4\pi}\Phi_s^-) \nn 
&-& \left. g_3  \cos(\sqrt{4\pi}\Theta_s^-)
\right\}. \label{Humk}
\eea
Within our $U$-$V$ model, the parameters are $g_1 = g_c = (U-3V)/n$, $g_3 = U/n$.
The slightly strange looking  structure
in the spin sector is an artifact of 
treating the non-Abelian $SU(2)$ symmetry group by means of 
Abelian bosonization; this symmetry becomes manifest upon refermionization (see next section).

Appendix \ref{fullinteraction} contains
a discussion of the full structure of the interaction including the less relevant terms which we will ignore in 
the remainder of this paper.  The only feature generated by this extra complexity is a minor logarithmic renormalization of
the masses generated in the theory.  This may be important for estimating quantities to be measured experimentally, 
but will not make a difference to any of the universal properties of the 
gapped ground state phases of the system and the quantum criticality separating them.


The staggered part of interaction (\ref{interaction})
is contributed both  
by the spin-exchange and non-spin-exchange scattering processes.
In the effective one-dimensional picture, the non-spin-exchange process is proportional to the sum
$
\sum_l (-1)^l n(l)n(l+1).
$
Here $n(l)$ is the total local density that transforms in the continuum limit
to
$
b [J(x)+(-1)^l M(x)],
$
where
$
J=\sum_{\mu\sigma}
\left(:R^{\dagger}_{\mu\sigma}R_{\mu\sigma}:  +  :L^{\dagger}_{\mu\sigma}L_{\mu\sigma}:\right)
$
and
$
M=\sum_{\mu\sigma}(R^\dagger_{\mu\sigma}L_{\mu\sigma} +h.c.).
$
The above sum then becomes
$
b \int \rd x~ [M,J]_{x,x+\alpha}.
$
The emerging point-split commutator is then estimated using Operator Product Expansion (see Appendix \ref{OPE}) and
is proportional to the single-particle dimerization operator (\ref{dimop2})
\bea
[M,J]_{x,x+\alpha} &\equiv& \lim_{\alpha \to 0}\left\{ M(x)J(x+\alpha) - J(x)M(x+\alpha) \right\}
\nonumber\\
&=& 
2{\cal O}_{\rm dim} (x).
\eea
A similar procedure results in the same operator for the spin-exchange processes.
Within the $U$-$V$ model  the dimerization amplitude is $g_{\Delta} = V/2n$.  We must stress again here that although ${\cal O}_{\dim}$ is a single particle dimerization operator, $g_\Delta$ is an {\it effective} coupling constant, arising from projecting the interactions on the hexagonal lattice onto the low-energy theory relevant for a zigzag nanotube.  The {\it bare} coupling constant associated with this operator arising from curvature effects may then be safely neglected, as it is of order $1/n^2$, much less than the effective $g_\Delta$.

Thus we arrive at the effective model describing the zigzag nanotube at half filling in the scaling limit,
\be
\ham = \ham_{\rm kin} + \ham_{\rm Coul} + \ham_{\rm umkl} + \ham_\Delta, \label{ham-eff-total}
\ee
where ${\cal H}_\Delta = g_\Delta {\cal O}_{\rm dim}$.

\section{Adiabatic approximation and mapping to spin sector}

The unscreened Coulomb interaction strongly enhances the velocity of the symmetric charge mode
relative to the group velocities of collective excitations in 
other channels.\cite{glaz92}
Hence, from the point of view of the field $\Phi_c^{+}$, all other degrees of freedom 
can be regarded as
static.  This allows one to 
employ an adiabatic approximation and  obtain the low-energy dynamics of the model by integrating out 
the total charge mode.
Such a procedure has already been 
discussed 
for narrow-gap SWNTs away from half-filling \cite{lt03} ($\ham_{\rm umkl}=0$) 
and for a half-filled armchair nanotube \cite{nt03}($\ham_\Delta = 0$).
The general analysis of the model
(\ref{ham-eff-total}) represents a more complicated task  
due to the presence of two
strongly relevant perturbations -- the Umklapp and dimerization terms,
each containing the field $\Phi_c^{+}$. 

In this paper, we will specialize to the limit where $\ham_{\rm umkl}$ is the 
principal term,
responsible for the formation of a fully gapped Mott phase, and then address the effect
of $\ham_\Delta$ as a perturbation on this phase.

Such an approach can be justified in
our $U$-$V$ model by choosing $U \gg V > 0$, because the structure of the Umklapp interaction (\ref{Humk}) contains parts proportional to $U$, whereas the dimerization includes only the smaller $V$.  The details of integrating out $\Phi_c^{+}$ are well explained in Ref. \onlinecite{nt03}.  Here we give a basic physical argument that leads us to the correct result.  
By examining Eq.(\ref{Humk}) and making the approximation that all other fields are static during the typical fluctuation time of $\Phi_c^{+}$, we obtain a modified sine-Gordon model describing the total charge field,
\be
{\cal H} \sim {\cal H}_0 [\Phi_c^{+}] - A \cos \sqrt{4\pi} \Phi_c^{+}. \label{modiSG}
\ee
Here ${\cal H}_0 [\Phi_c^{+}]$ is the sum of a standard Gaussian model and the nonlocal Coulomb term 
(\ref{lr-coul}),
whereas $A$ is a combination of the slower fields which, within the adiabatic approximation, can be
replaced by a constant.
Due to the high velocity and low effective scaling dimension of the field $\Phi_c^{+}$, all 
the spectral gaps (i.e. those of solitons and breathers) 
in the total charge sector 
are very large.  Therefore, to examine the low-energy dynamics of 
neutral modes, one can consider the field $\Phi_c^{+}$ to be locked at $\Phi_c^{+}=0$
and replace cosines of this field by their expectation values.

It is important at this stage to consider the physics we are discarding by making this approximation.  Charged solitons in the $\Phi_c^+$ field do indeed exist in the theory.  Furthermore, electron like quasi-particles described by a half soliton simultaneously in each of the sectors of the theory are also present. Finally, there will also be plasmonic excitations corresponding to breathers modes in Hamiltonian (\ref{modiSG}).  However, due to the strong renormalization of the $\Phi_c^+$ field, each of these excitations will have a large mass.

Thus, as we are interested in the low-energy properties of the model, the only excitations that remain are neutral collective modes: the relative charge ``vortex''\cite{nt03} modes, and the spin degrees of freedom (which as we will see later can be separated into a singlet mode and a triplet mode), and it is in this region of energies below any of the charged excitations that the adiabatic approximation (and therefore the description in terms only of neutral modes) is valid.  This also explains why the dimerization term ${\cal H}_\Delta$ with a magnitude much less than any of the charged excitations can not turn the system into a trivial band-insulator: everything occurs within the Mott phase.  We now proceed to analyse the low-energy phase diagram of such a model, and show the existence of an interesting non-trivial quantum phase transition.

\subsection{The Umklapp term}

Following the preocedure of Ref. \onlinecite{nt03}, one can identify the remaining neutral collective modes at 
$g_\Delta = 0$ 
by 
introducing four real (Majorana) fermions $\chi^i _{R,L}$ $(i=0,1,2,3)$ and
refermionizing the theory.
The resulting Hamiltonian displays an SU(2) $\times$ Z$_2$ symmetry in the spin sector:
\bea
\ham' &=& \frac{v}{2}\left[ \left(\p_x \Theta^- _{c} \right)^2  + \left(\p_x \Phi^- _{c} \right)^2\right]
- \frac{m_f}{\pi\alpha} \cos \sqrt{4\pi} \Phi^- _{c} \nn
&+& \frac{iv}{2} 
\sum_{i=0}^3 (-\chi_R^i \partial_x \chi_R ^i
+ \chi_L^i \partial_x \chi_L^i ) 
-im_t\sum_{a=1}^3\chi_R^a\chi_L^a \nn
&-& im_s \chi_R^0\chi_L^0. \label{refermionized}
\eea
Here 
$m_f = \langle \cos(\sqrt{4\pi}\Phi^+ _c) \rangle g_1/\pi\alpha$ is the mass of the relative charge 
(``vortex'') excitation, whereas 
$m_t=\langle \cos(\sqrt{4\pi}\Phi^+ _c) \rangle(g_1-g_3)/\pi\alpha$ 
and $m_s=\langle \cos(\sqrt{4\pi}\Phi_c^+) \rangle(g_3+g_1)/\pi\alpha$
are the masses of the spin-triplet and spin-singlet modes, respectively.  In terms of our 
$U-V$ model, the masses are 
parametrized as follows: 
\be
m_s = C \frac{ 2U-3V}{n}, \;\; m_t = -C \frac{3V}{n},
\ee
where $C= \langle \cos(\sqrt{4\pi}\Phi^+ _c) \rangle \approx 1$.

Associated with the  Majorana fermions are four Ising models characterized by order and disorder parameters, $\sigma_i$ and $\mu_i$.  We use the correspondence:
\bea
\cos \sqrt{\pi}\Phi_s^+ = \sigma_1\sigma_2 & \;\;\; &
\sin \sqrt{\pi}\Phi_s^+ = \mu_1\mu_2 \nn
\cos \sqrt{\pi}\Theta_s^+ = \mu_1\sigma_2 & \;\;\; &
\sin \sqrt{\pi}\Theta_s^+ = \sigma_1\mu_2 \nn
\cos \sqrt{\pi}\Phi_s^- = \sigma_0\sigma_3 & \;\;\; &
\sin \sqrt{\pi}\Phi_s^- = \mu_0\mu_3 \nn
\cos \sqrt{\pi}\Theta_s^- = \mu_0\sigma_3 & \;\;\; &
\sin \sqrt{\pi}\Theta_s^- = \sigma_0\mu_3.
\eea
 The sign of the Majorana mass indicates whether the
corresponding Ising model is ordered ($m <0$, $\la  \sigma  \ra \neq 0$, $\la  \mu  \ra = 0$) or
disordered ($m >0$, $\la  \mu  \ra \neq 0$, $\la  \sigma  \ra = 0$) \cite{bosonization}.

  The fermionic part of the Hamiltonian (\ref{refermionized}) has the same structure as that
of a two-leg antiferromagnetic spin-1/2 ladder.\cite{snt96} Indeed,
assuming that $U/V \gg 1$, we find that $m_t<0$ and $m_s>0$ so that 
$\langle \sigma_a \rangle \ne 0$ for $a=1,2,3$ and 
$\langle \mu_0 \rangle \ne 0$.  
Furthermore, $m_f>0$ meaning that 
$\langle \cos(\sqrt\pi \Phi^- _c) \rangle \ne 0$. These signs of the masses indicate 
a spin-liquid behavior of the system.\cite{snt96,bosonization}  We can further quantify this by defining the staggered magnetization as a
suitably averaged difference between the local
spin densities on the two sublattices, 
${\bf n}^- ({\bf r}) = {\bf S}_a ({\bf r}) -{\bf S}_b ({\bf r})$.
Projecting this onto the low-energy sector of the model gives
\be
{\bf n}^- \sim \cos(\sqrt{\pi} \Phi_{c^-})  \mu_0 \,
(\mu_1\sigma_2\sigma_3,\sigma_1\mu_2\sigma_3,\sigma_1\sigma_2\mu_3 ).
\label{stag-mag}
\ee
The two-point correlation function of ${\bf n}^-$ displays a coherent magnon peak
with a mass gap $|m_t|$.

\subsection{Effect of dimerization term}

After refermionization the  dimerization operator acquires the following low-energy form:
\be
{\cal O}_{\dim} \sim \cos(\sqrt\pi \Phi^- _c) \mu_0 \mu_1 \mu_2 \mu_3.\label{dimreferm}
\ee
In the spin-liquid phase formed by the umklapp processes this operator can be further simplified
by replacing in the leading order the operators $\cos(\sqrt\pi \Phi_c^-)$ and $\mu_0$ by
their expectation values. Thus, projecting ${\cal O}_{\dim}$ onto the spin-triplet sector
yields
\bea
\ham_\Delta &=& h_{\reff} ~ \mu_1 \mu_2 \mu_3, \nn
h_{\reff} &=& \frac{4g_\Delta}{(\pi \alpha)^2}
\langle \cos(\sqrt{\pi}\Phi_c^+)\rangle  \langle \cos(\sqrt{\pi}\Phi_c^-)\rangle
\langle \mu_0 \rangle. \label{heff}
\eea
We see that the dimerization term competes with umklapp processes which support
the ground state with $ \la \sigma_i \ra \ne 0 $, $ \la \mu_i \ra = 0$.

The resulting theory 
thus involves only the spin-triplet degrees of freedom and actually coincides with
the problem of a two-leg spin ladder with explicit dimerization,\cite{mss96} or, equivalently, a 
weakly dimerized spin-1 chain with a small Haldane gap.\cite{wn00} It was shown that
increasing the dimerization can drive the system towards a quantum criticality
belonging to the universality class of the SU(2)$_1$ Wess-Zumino-Novikov-Witten model
with central charge $1$. Close to criticality the universal 
lowest-energy properties of the system are those
of a single, explicitly dimerized, antiferromagnetic spin-1/2 chain,
 with the dimerization changing its sign across the transition.

The location of the critical point can be estimated by requiring that the mass gap generated by the
dimerization term alone and the triplet mass become of the same order.
Using standard scaling arguments we find that
in Eq. (\ref{heff})
$
h_\reff \sim |m_f|^{1/4} |m_s|^{1/8}
g_\Delta,
$
where all parameters in the right-hand side are proportional to $1/n$.
Therefore the dimerization gap scales as
\be
m_{dim} \sim h_\reff^{8/13} \propto (1/n)^{11/13},
\ee 
and so, within our $U \gg V$ approximation
\be
\frac{m_\Delta}{|m_t|} \propto \left(\frac{\bar{U}}{\bar{V}}\right)^{3/13} 
\left(\frac{n}{\bar{V}}\right)^{2/13}, 
\ee
where $\bar{U} = U \alpha/v$ and $\bar{V} = U \alpha/v$ are dimensionless coupling constants.

This estimate shows that for sufficiently large radius $n$ the dimerization is always dominant 
and 
the system occurs in a gapped phase 
whose properties are governed by the operator $O_{dim}$.  Whether a crossover 
to the spin-liquid phase through the SU(2)$_1$ criticality
can occur as a function of the nanotube's radius depends on various nonuniversal prefactors.  
As shown in Appendix~\ref{fullinteraction}, an improved estimate of the critical line
can be obtained by taking less relevant (non-umklapp) terms into account. The ratio of the dimerization
and triplet mass gaps then becomes
\be
\frac{m_{dim}}{|m_t|} \sim \frac{(\bar{V}/\bar{U})^{8/13}}{(\bar{U}/n)^{15/13} \ln (n/\bar{U})}.
\label{newratio}
\ee
The new estimate shows that the quantum transition scenario 
taking place upon decreasing $n$ at large $U$ 
is a real possibility.

\begin{figure}
\begin{center}
\epsfig{file=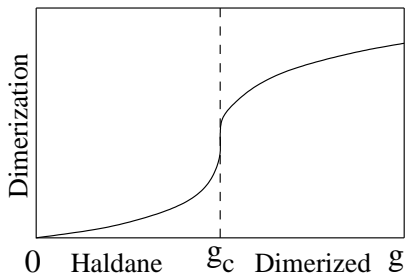,width=2.1in}
\caption{Average dimerization as a function  of $g=m_\Delta/m_t\sim n^{2/13}$,
being nonzero in both gapped phases, displays a singularity in the derivative
at the critical point $g = g_c$.}
\label{qcpfig}
\end{center}
\end{figure}

\begin{figure}
\begin{center}
\epsfig{file=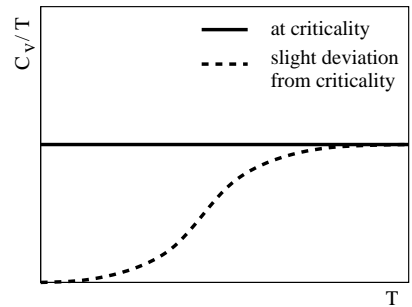,width=2.1in}
\caption{$C_v/T$ as a function of $T$.  The solid line denotes the behavior exactly at criticality $g=g_c$, the dashed line is slightly away from criticality.  The crossover temperature is given by the smallest gap in the system $m\sim|g-g_c|^{2/3}$.}
\label{thermo}
\end{center}
\end{figure}

\subsection{Physical Observables}

The ratio $g=m_\Delta/m_t$ acts as a control parameter for the quantum phase transition, with a critical point $g_c \sim 1$.  As we vary $g$, the dimerization itself shows a singularity in its derivative as one passes through the transition point itself\cite{wn00} according to
\be
\la {\cal O}_{dim} \ra = \la O_{dim} (g=g_c) \ra + \alpha |g - g_c|^{1/3} {\rm sign} (g - g_c),
\ee
where $\alpha$ is some constant. This is qualitatively shown in Fig. \ref{qcpfig}.

The parameter $g$ can be varied by varying the radius of the nanotube, $n$, or by changing the interaction coupling constants; the latter can be achieved by, for example, applying pressure or 
stretching the nanotube.  The dimerization of the nanotube will show up as a Peierls distortion of 
the lattice, which should be measurable with low-temperature STM.

The proximity to the quantum critical point will also show up in thermodynamic quantities, such as specific heat.  Exactly at the critical point, the specific heat is linear in temperature\cite{bcn}
\be
C_V(g=g_c) = c \frac{\pi T }{3 v},
\ee
where $c=1$ is the central charge of the criticality, and $v$ is the Fermi velocity.  Away from criticality, the specific heat is exponentially small at sufficiently low temperature $T<m$,
\be
C_V \sim T^{-3/2} e^{-m/T},
\ee
where $m$ is the smallest gap in the system, which near criticality is given\cite{wn00} by $m\sim| g-g_c|^{2/3}$.  The specific heat crosses over to a linear dependence on temperature when $T>m$.
This is plotted in Figure \ref{thermo}.

\section{Conclusions}

We have shown that in zigzag carbon nanotubes electron-electron interactions play an
important role in the ground state of the system.  Firstly, there is the long-range
unscreened Coulomb interaction, which sets a hierarchy of energy scales in the problem.  Secondly, there are umklapp processes (in the undoped case) which gap all collective excitations in the nanotube.  Finally, there is an explicit dimerization in the interaction, scaling as $1/n$ due to the way that we break the $C_3$ symmetry when we wrap the zigzag nanotube.  This dimerization originates from interactions
and affects the effective low-energy action; there it directly competes with Umklapp processes
supporting a Mott-like insulating phase with 
a spin-liquid structure of collective excitations.

As the relative strength of the dimerization is increased, the system can exhibit a nontrivial quantum criticality with an $SU(2)_1$ symmetry.  On one side of this critical point, the system is in a Haldane spin liquid phase, on the other side of the critical point, the ground state is dimerized.  The position in the phase diagram depends on the radius of the nanotube, with reasonable assumptions about parameters showing we are always near the QCP.  Therefore, small external perturbations may be able to drive the real system of a zigzag carbon nanotube to an exciting theoretical quantum criticality.

\section{Acknowledgements}

We thank A.M.Tsvelik and F. Essler for stimulating discussions.  STC's work is supported by the EPSRC grant GLGL RRAH 11382, and was partly done while STC was at ICTP.
Part of this work was done during AAN's visit to
Imperial College supported by EPSRC grant GR/N19359.
AOG's work is partly supported by the EPSRC grant GR/R70309 by the EU training network DIENOW.

\appendix

\section{Full structure of interaction}
\label{fullinteraction}

The long range Coulomb interaction means that the gap in the total charge
excitations is the largest energy scale in the problem, and consequently the umklapp terms which involve $\cos\sqrt{4\pi}\Phi_c^+$ are the most relevant terms which gap the remaining sectors of the theory.  However, as the term which gaps the triplet sector is proportional to $V \ll U$, less relevant terms may strongly renormalize the triplet mass.  In this section, we continue to assume that $U \gg V$, so that we only look at corrections proportional to $U$.

Bosonizing the interaction $H_U$ (\ref{interaction}) gives rise to the following addition to the Umklapp Hamiltonian (\ref{Humk})
\be
\Delta H = \bar{U} \int dx
\cos \sqrt{4\pi} \Phi_c^- \left( \cos\sqrt{4\pi} \Phi_s^+ + \cos\sqrt{4\pi} \Theta_s^- \right),
\ee
where $\bar{U}$ is a dimensionless interaction constant $\bar{U}=Ub/(\pi\alpha)^2$.

The mass gap (in dimensionless units) associated with the relative charge sector is generated by the more relevant 
umklapp processes
\be
m_f \approx C \frac{\bar{U}}{n}
\ee
where $C=\la \cos \sqrt{4\pi} \Phi_c^+ \ra \approx 1$, so we can assume that this field $\Phi_c^-$ becomes locked, and the expectation value
\be
\la \cos\sqrt{4\pi} \Phi_c^- \ra \approx m_f \ln\frac{1}{m_f}.
\ee

After refermionization, $\Delta H$ will contribute to both the singlet mass and the triplet mass.  As the singlet gap is already of order $U/n$, a change of order $(U/n)^2$ is insignificant.  However, the triplet mass will become
\be
m_t = -C \frac{3\bar{V}}{n} - C^2 \left(\frac{\bar{U}}{n}\right)^2 \ln \left( \frac{n}{\bar{U}} \right).
\ee
We considered in the main text the case where $V/n \gg (U/n)^2 \ln (n/U)$.  If $U$ is large enough, then the opposite limit may hold true, in which case the ratio of the dimerization and the triplet 
masses modifies and becomes given by formula (\ref{newratio}).

\section{The Operator Product Expansion}
\label{OPE}

Here, we use the Operator Product Expansion (OPE) in the fermionic basis to show that the staggered interaction gives rise to a single particle dimerization term in the continuous limit.

The bare Hamiltonian in the Fermionic basis is
\be
H_0 = -i v_F \sum_{\mu\sigma} \int dx \left[ R^\dagger_{\mu\sigma} \p_x R_{\mu\sigma} - L^\dagger_{\mu\sigma} \p_x L_{\mu\sigma} \right],
\ee
so the Matsubara Green's functions are
\bea
\la L_{\mu\sigma}^\dagger (z) L_{\mu'\sigma'} (w) \ra = 
\la L_{\mu\sigma} (z) L_{\mu'\sigma'}^\dagger (w) \ra &=&
\frac{\delta_{\mu\mu'}\delta_{\sigma\sigma'}}{2\pi(z-w)} \nn
\la R_{\mu\sigma}^\dagger (\bar{z}) R_{\mu'\sigma'} (\bar{w}) \ra = 
\la R_{\mu\sigma} (\bar{z}) R_{\mu'\sigma'}^\dagger (\bar{w}) \ra &=&
\frac{\delta_{\mu\mu'}\delta_{\sigma\sigma'}}{2\pi(\bar{z}-\bar{w})} \nn
\eea
where
\be
z = v_F \tau + ix, \; \; \bar{z} = v_F \tau - ix,
\ee
and similarly for $w$, are the complex coordinates written in a manifestly Lorentz invarient way, with $\tau$ being the imaginary time.

The first staggered piece in Eq. (\ref{interaction}) takes the form
\be
(-1)^l n(l) n(l+1),
\ee
where $n(l)=\sum_{\mu\sigma} n_{\mu\sigma} (l)$ is the total electron density at site $l$.  In the continuum limit, this becomes
\be
b \left[ M(x) J(x+\alpha) - J(x) M(x+\alpha) \right], \label{fst}
\ee
where
\bea
J(x) &=& \sum_{\mu\sigma} \left( J^R_{\mu\sigma} + J^L_{\mu\sigma} \right), \nn
M(x) &=& \sum_{\mu\sigma} \left( R^\dagger_{\mu\sigma} L_{\mu\sigma} + L^\dagger_{\mu\sigma} R_{\mu\sigma}\right).
\eea
Therefore, in the expansion of (\ref{fst}), we will need the following OPE:
\bea
&& : L^\dagger (z) L(z) : : L^\dagger(w) R(w) : \nn &\sim& \la L(z) L^\dagger(w) \ra
 : L^\dagger(z) R (z) : = \frac{1}{2\pi(z-w)} : L^\dagger(z) R (z) :, \nn
&& : L^\dagger (z) L(z) : : R^\dagger(w) L(w) : \nn &\sim& -\la L^\dagger( z) L (w) \ra
 : R^\dagger(z) L (z) : = \frac{-1}{2\pi(z-w)} : R^\dagger(z) L (z) :, \nn
\eea
where the flavor and spin subscripts must all be identical.  Similar OPE's hold for the
antianalytic part of the expansion, so that
\bea
J(z) M(w) &\sim& 
 \frac{1}{2\pi} \left( \frac{1}{z-w} - \frac{1}{\bar{z}-\bar{w}} \right)
\nn && \times \sum_{\mu\sigma}:\left( R^\dagger_{\mu\sigma} L_{\mu\sigma} - L^\dagger_{\mu\sigma} R_{\mu\sigma}\right) :.
\eea
We recognise the operator as the dimerization operator, hence putting $\tau=0$ and evaluating the point-split commutator in (\ref{fst}), we see that
\bea
[M,J]_{x,x+\alpha} &\equiv& \lim_{\alpha \to 0}\left\{ M(x)J(x+\alpha) - J(x)M(x+\alpha) \right\}
\nonumber\\
&=& 
2O_{\rm dim} (x).
\eea

\begin{widetext}
The other component of the staggered interaction
\bea
&& (-1)^l c_{\mu\sigma}^\dagger(l) c_{-\mu\sigma}(l) c^\dagger_{-\mu\sigma'}(l+1) c_{\mu\sigma'}(l+1) \nn
&=& b^2\left[
\left(R^\dagger_{\mu\sigma} L_{-\mu\sigma} + L^\dagger_{\mu\sigma} R_{-\mu\sigma} \right)_x
\left(R^\dagger_{-\mu\sigma'} R_{\mu\sigma'} + L^\dagger_{-\mu\sigma'} L_{\mu\sigma'} \right)_{x+\alpha} \right. \nn
&-& \left.
 \left(R^\dagger_{\mu\sigma} R_{-\mu\sigma} + L^\dagger_{\mu\sigma} L_{-\mu\sigma} \right)_x
\left(R^\dagger_{-\mu\sigma'} L_{\mu\sigma'} + L^\dagger_{-\mu\sigma'} R_{\mu\sigma'} \right)_{x+\alpha} \right]
\eea
can be treated in an identical way.  This too turns into the dimerization operator, giving the final result quoted in the main text.
\end{widetext}


\begin{thebibliography}{16}
\bibitem{footnote}
In the non-interacting case, $b=\pi\alpha$.  In the presence of interactions, this relationship changes in a non-universal way, however the two parameters stay within the same order of magnitude and the difference between them is not important for the purposes of this paper.
\end{thebibliography}
\end{document}